\documentclass[]{spie}
\title{Photonic lantern wavefront reconstruction in a multi-wavefront sensor single-conjugate adaptive optics system}

\author[a]{Aditya R. Sengupta}
\author[a]{Jordan Diaz}
\author[b]{Benjamin L. Gerard}
\author[a]{Rebecca Jensen-Clem}
\author[a]{Daren Dillon}
\author[a]{Matthew DeMartino}
\author[a]{Kevin Bundy}
\author[c]{Sylvain Cetre}
\author[a]{Vincent Chambouleyron}

\affil[a]{Department of Astronomy \& Astrophysics, University of California, Santa Cruz, CA 95064, USA}
\affil[b]{Lawrence Livermore National Laboratory}
\affil[c]{Wakea Consulting}

\authorinfo{Further author information: (Send correspondence to A.S.)\\A.S.: E-mail: adityars@ucsc.edu}

\usepackage{preamble}
\begin{document} 
\maketitle

\begin{abstract}
Exoplanet direct imaging using adaptive optics (AO) is often limited by non-common path aberrations (NCPAs) and aberrations that are invisible to traditional pupil-plane wavefront sensors (WFSs). This can be remedied by focal-plane (FP) WFSs that characterize aberrations directly from a final science image. Photonic lanterns (PLs) can act as low-order FPWFSs with the ability to direct some light to downstream science instruments. Using a PL on the SEAL (Santa Cruz Extreme AO Laboratory) high-contrast imaging testbed, we demonstrate (1) linear ranges and (2) closed-loop control. Additionally, we simulate the use of the PL in a multi-wavefront sensor AO system, in which multiple WFSs feed back to the same common-path deformable mirror. Building on previous multi-WFS AO demonstrations on SEAL, we simulate a modulated pyramid WFS to sense aberrations of high spatial order and large amplitude, and the PL to sense low order aberrations including NCPAs. We assess adaptive optics performance in this setting using three different PL wavefront reconstruction algorithms. We also provide a new method to experimentally identify the propagation matrix of a PL, making advanced model-based algorithms practical. This work demonstrates the role of photonic technologies and multi-stage wavefront sensing in the context of extreme AO and high contrast imaging.
\end{abstract}

\keywords{Astrophotonics, photonic lantern, wavefront sensing, wavefront control, adaptive optics, multi-stage adaptive optics}

\section{INTRODUCTION}

High-contrast adaptive optics (AO) systems often benefit from wavefront sensing and control on the final science image, which can correct optical aberrations that appear only in the focal plane and are not seen by pupil-plane wavefront sensors. The photonic lantern (PL)\cite{Norris20} is particularly suitable for this task, as it allows focal-plane light to be used for wavefront sensing while some or all of it is directed to a science instrument downstream\cite{Norris22}. PLs consist of a multi-mode optical fiber input that tapers to several single-mode fiber outputs. The state of the wavefront is encoded in the distribution of intensities across the output fibers. 

On their own, PLs are not able to sense and reconstruct the full scale of turbulence seen at most major observatories; the number of modes of optical aberrations they are able to reconstruct is limited by their number of output ports (we refer to this as $N$ and take $N = 19$ in this work), and the dynamic ranges achieved by real lanterns are not yet well characterized. There is therefore a need for the PL to operate together with a first-stage pupil-plane wavefront sensor in order to achieve optimal correction. In this work, we present AO system simulations using the PL in a multi-WFS single-conjugate AO (SCAO) configuration, first introduced by Gerard et al. (2021)\cite{BenSecondStage2021} and implemented on the Santa Cruz Extreme AO Laboratory (SEAL)\cite{SEAL} testbed using the FAST FPWFS by Gerard et al. (2023)\cite{BenMultiWFS}. This mode is characterized by multiple wavefront sensors feeding back to a single common-path deformable mirror (DM).

The optimal technique for reconstructing wavefronts from PL outputs has yet to be determined, and this choice is likely to affect AO system performance. Therefore, we provide simulation results using three different reconstruction techniques: linear reconstruction using the traditional AO method of measuring an interaction matrix, a modified version of the neural network method introduced in Norris et al. (2020)\cite{Norris20} used to assess the potential of data-driven methods, and an implementation of the Gerchberg-Saxton algorithm used to assess the potential of model-driven methods. We test linear reconstruction on SEAL and present linearity curves. We also present a new method to exactly find the propagation matrix of a PL in $N^2$ measurements, making model-driven wavefront reconstruction methods directly applicable in experimental settings.

The remainder of this paper is structured as follows. Section~\ref{sec:lantcontrol} discusses the methods used for simulating photonic lanterns within AO systems. Section~\ref{sec:reconstruction} discusses the reconstruction algorithms in detail and assesses their performance, along with the empirical characterization algorithm. Section~\ref{sec:results} discusses the multi-WFS SCAO architecture employed and presents the results of the corresponding simulations. Section~\ref{sec:conclusion} is a conclusion.  

\section{LANTERN SIMULATION}
\label{sec:lantcontrol}

\subsection{Photonic lantern simulation}

We simulate photonic lanterns using the \textit{lightbeam} Python package\cite{lightbeam}, which implements the finite-difference beam propagation method (FD-BPM) for weakly guiding waveguides. In this work, we use a 19-port photonic lantern operating at $\lambda = 1.55 \mu$m. The cladding refractive index is 1.4504 and the core refractive index is 1.46076. The cladding diameter at the entrance is 37 $\mu$m, the core diameter is 4.4 $\mu$m, and the cores are offset from the center by 7.4 $\mu$m for the inner ring and 14.8 $\mu$m for the outer ring. The lantern is 60 mm long and has a tapering factor of 8, i.e. the dimensions at the output are the same as those at the input multiplied by 8. \textit{lightbeam} runs are carried out with a grid resolution of 1 $\mu$m in the $x$ and $y$ directions, 50 $\mu$m in the $z$ direction, an extent in $x$ and $y$ of 512 $\mu$m, and 8 grid cells on each boundary in $x$ and $y$ used as perfectly matched layers.

To simulate the response of the photonic lantern to arbitrary input electric fields without needing to run beam propagation each time, we follow the method of Lin et al. (2022)\cite{Lin22}, in which each of the output single-mode fiber ports is illuminated with the fundamental mode (LP${}_{01}$) and the electric field at the input that results from numerically back-propagating from this state is recorded, forming a basis of input electric fields. Due to the linearity of the photonic lantern in electric field, we can project any input onto this basis, and the coefficients of this projection correspond to the reading at the PL output; specifically, the coefficient on the $i$th basis element is the complex amplitude on the fundamental mode at the $i$th output port. The measurement taken by the photonic lantern is taken to be the norm-squared of these complex amplitudes. 

\subsection{Adaptive optics simulation incorporating the photonic lantern}

We simulate an adaptive optics loop using the \textit{hcipy}\cite{hcipy} Python package. We set the focal-plane resolution and extent to match the \textit{lightbeam} grids to avoid needing to resample. To simulate propagation through the lantern, FP electric fields are unraveled into 1D vectors over only the region covered by the PL input, and these are projected onto a basis of similar 1D vectors created using \textit{lightbeam} as detailed above. The coefficients of this projection are used directly, or embedded into an \textit{hcipy.Wavefront} object when necessary, such as for the Gerchberg-Saxton algorithm (\ref{sec:GS}).

We use a 20$\times$20-actuator deformable mirror with influence functions defined in modal space over the Zernike modes. We use a modulated pyramid wavefront sensor for correction before the photonic lantern. We set the modulation radius to $5\lambda/D$ and use 12 steps per modulation cycle. We use an $f/6.5$ beam, resulting in a beam size of 10.075 $\mu$m, as this resulted in optimal wavefront sensing performance. Future work will address optimizing over beam size in more detail. We simulate turbulence with the \textit{hcipy.InfiniteAtmosphericLayer} object, with an outer scale of 50m and wind velocity of 10m/s. We allow $D/r_0$ to vary while keeping $D$ fixed to 10m.

We measure AO performance by measuring the Strehl ratio on the PSF as injected into the photonic lantern, using the \textit{get\_strehl\_from\_focal} function in \textit{hcipy}.

\section{WAVEFRONT RECONSTRUCTION}
\label{sec:reconstruction}

The performance of the AO system in its dual-WFS configuration may be significantly affected by the choice of reconstruction algorithm used by the photonic lantern. The response of the photonic lantern as measured in intensity is non-linear, and the combination of residual wavefront errors and non-common-path aberrations may be sufficiently large that linear reconstruction becomes inaccurate. To address this, we compare three reconstruction algorithms: linear reconstruction, neural network reconstruction as in Norris et al. (2020)\cite{Norris20}, and the Gerchberg-Saxton phase retrieval algorithm. In all cases, we reconstruct 9 Zernike modes: x-tilt, y-tilt, focus, two modes of astigmatism, and four modes of coma\footnote{following the \href{https://mthamilton.ucolick.org/techdocs/instruments/ShaneAO/gallery/}{ShaneAO convention}.}. The 19-port PL is inherently restricted to correcting 18 modes; we make a further restriction in order to minimize modal crosstalk while still capturing the dominant components of residual optical aberrations after the pyramid.

We present results from linear reconstruction on SEAL, and from all three methods in simulation.

\subsection{Linear reconstruction}
\label{sec:linear}

\begin{figure}[!htbp]
    \centering
    \includegraphics[width=\textwidth]{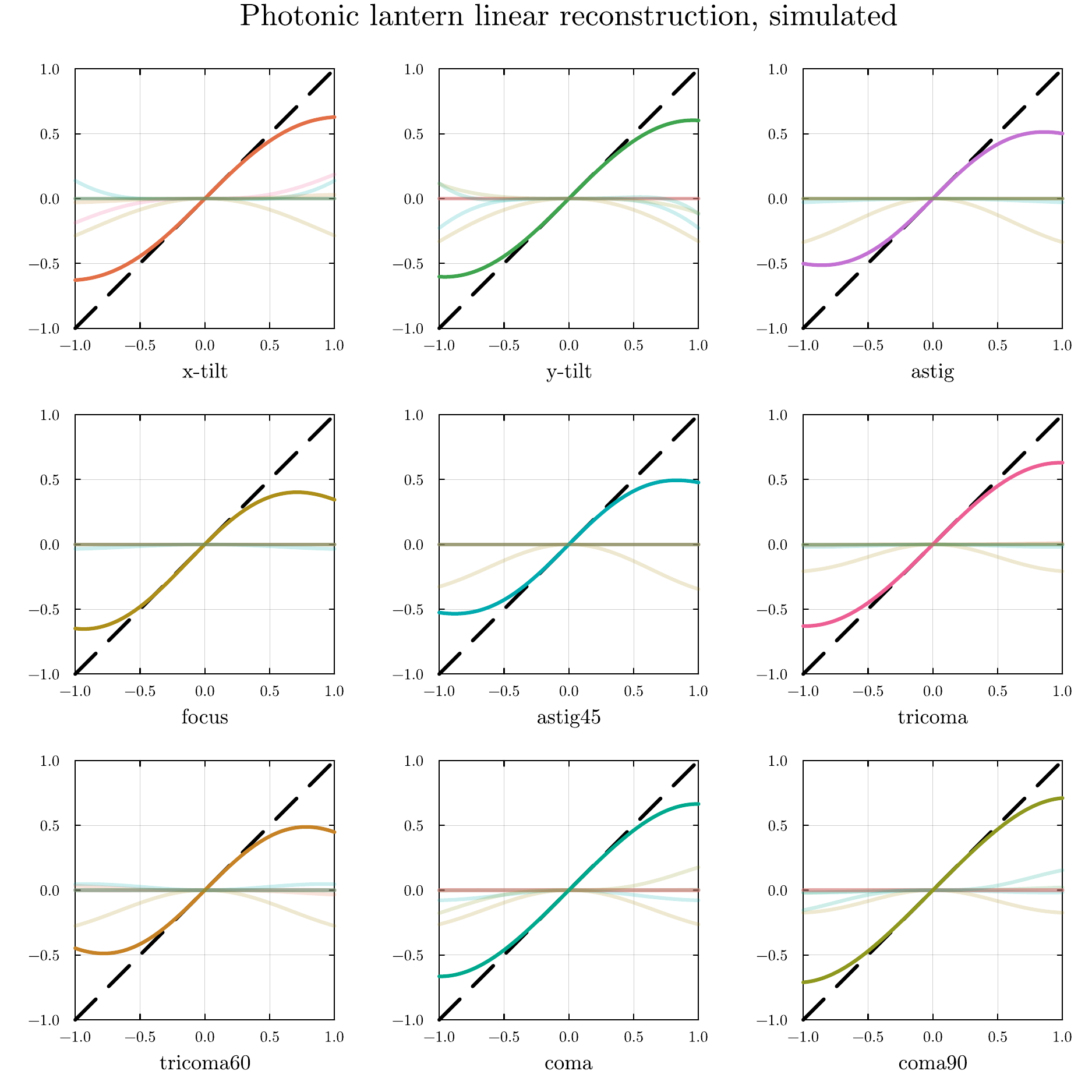}
    \caption{Simulated linearity curves with the linear reconstructor (injected vs. recovered signal, both in radians). The high-opacity line is the injected mode and the low-opacity lines are the other reconstructed modes.}
    \label{fig:linear_sim}
\end{figure}

We follow the standard method for linear reconstruction of a WFS signal. We compute an interaction matrix by recording the response of the PL to a small positive and small negative perturbation in each Zernike mode, allowing us to compute slopes around a flat position. The resulting matrix is inverted to create a command matrix, and multiplying a PL reading (minus the flat reading) by this command matrix provides the best linear reconstruction of the injected phase. Linearity curves from this method are shown in Figure~\ref{fig:linear_sim}. Linear reconstruction using a basis of phase screens better suited to the lantern, which can be found via singular value decomposition of the interaction matrix (Figure~\ref{fig:diagonal_phase_screens}), may yield better results, but this is not considered here in the interest of more straightforward comparisons between the methods.

\begin{figure}[!htbp]
    \centering
    \includegraphics[width=\textwidth]{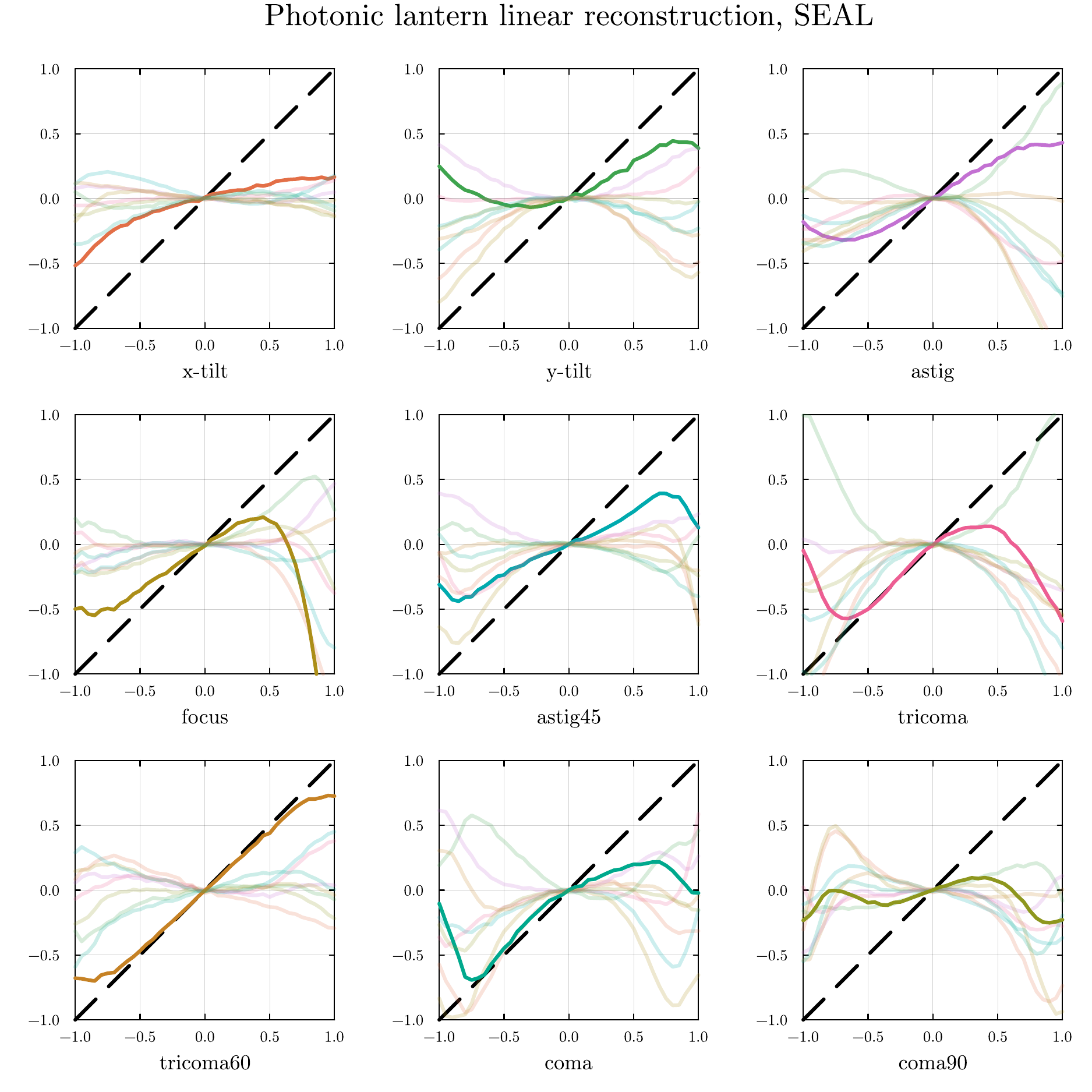}
    \caption{Linearity curves measured experimentally with the photonic lantern on SEAL. We observe significantly more modal cross-talk and smaller dynamic ranges.}
    \label{fig:linear_seal}
\end{figure}

\begin{figure}[!htbp]
    \centering
    \includegraphics[width=\textwidth]{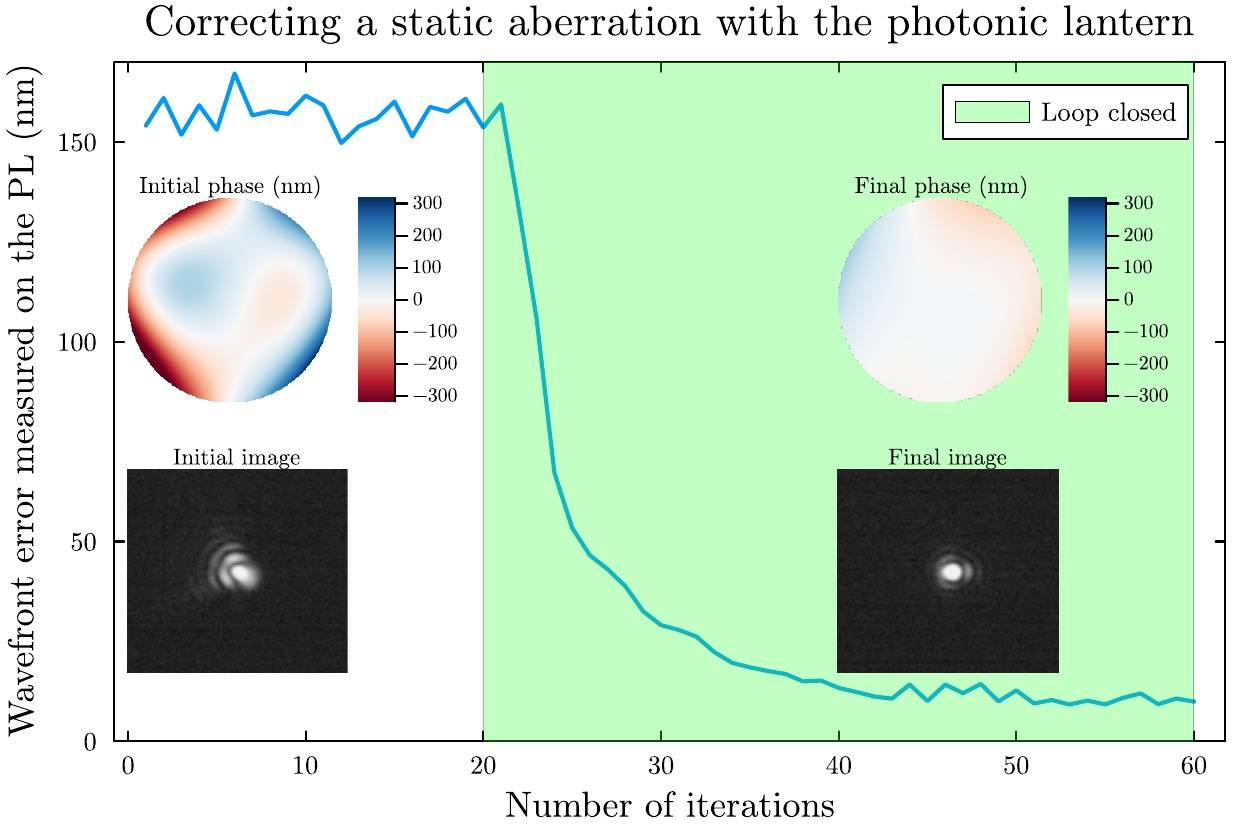}
    \caption{Correcting a static phase screen using the photonic lantern on SEAL.}
    \label{fig:seal_static_cl}
\end{figure}

We successfully implemented linear reconstruction with a photonic lantern on the SEAL testbed. Figure~\ref{fig:linear_seal} shows the measured linearity curves, and Figure~\ref{fig:seal_static_cl} shows a representative test of closed-loop control on a static aberration. Performance was likely to have been limited by a mismatch between the lantern's design wavelength (1550 nm) and the SEAL operating wavelength (635 nm).

Unfortunately, our PL broke in March 2024 and the replacement we procured broke in April 2024, making it impossible to experimentally confirm further simulation results. Future work will experimentally investigate our results regarding the control architecture, wavefront reconstruction, and AO performance. We will assess wavefront reconstruction quality via comparisons with SEAL's other wavefront sensors and making use of known wavefront aberrations introduced with the spatial light modulator\cite{MaaikeSLM}, and we will assess AO performance by placing a beam splitter in the focal plane before the photonic lantern and measuring the Strehl ratio on the resulting PSF image.

\subsection{Neural network reconstruction}

\begin{figure}[!htbp]
    \centering
    \includegraphics[width=\textwidth]{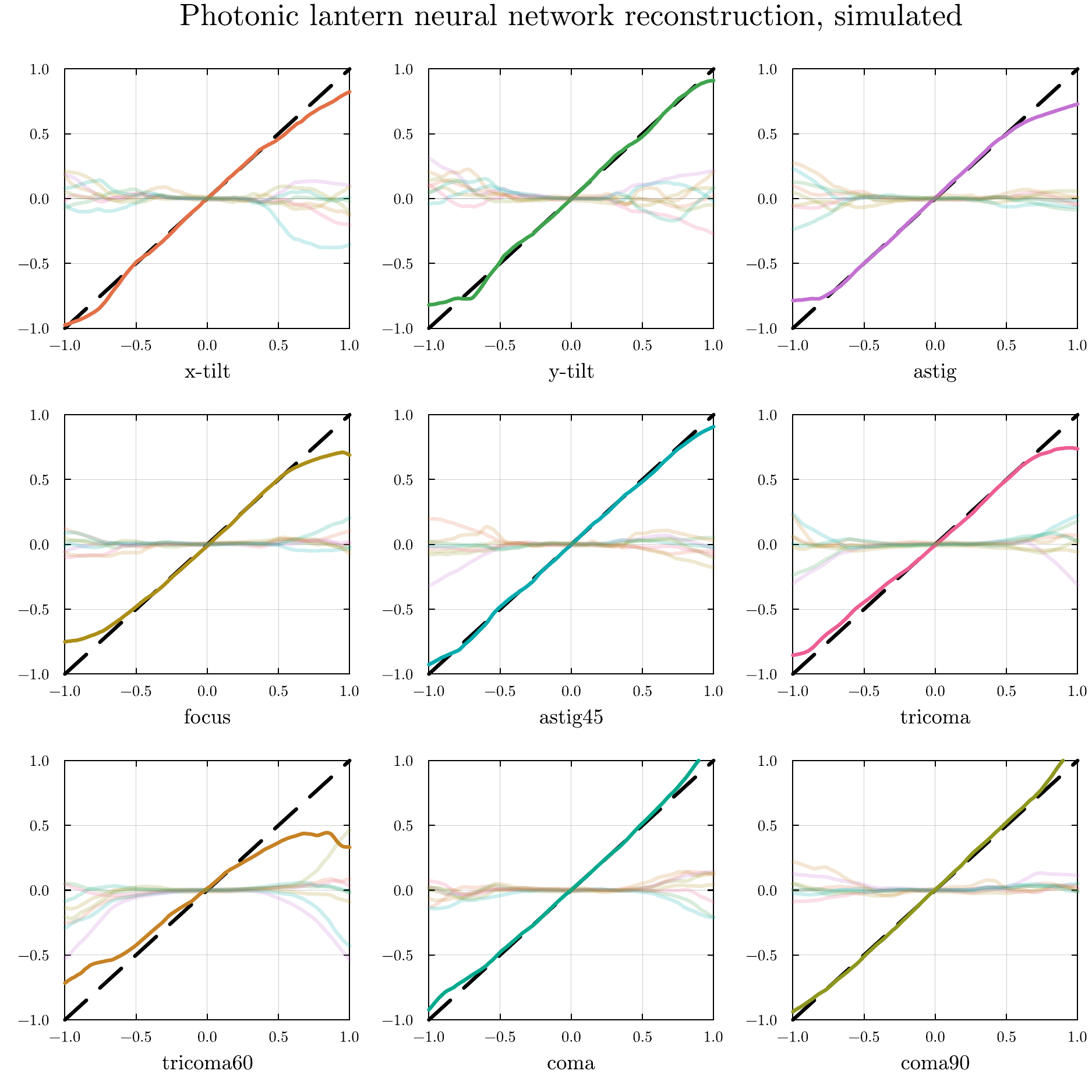}
    \caption{Linearity curves from the neural-network reconstructor (injected vs. recovered signal, both in radians). Modal crosstalk is notably reduced compared to the linear reconstructor.}
    \label{fig:nn_sim}
\end{figure}

We reproduce the best-performing neural network reconstructor from Norris et al. (2020)\cite{Norris20}. This model uses two hidden layers with 2000 and 100 neurons respectively, and uses a ReLU activation function. We train using an Adam optimizer on 48,000 measurements and use a test set of 12,000 measurements. We choose the training set slightly differently: all input phase screens are chosen by drawing a coefficient for each of the first 9 Zernike modes uniformly at random, then rescaling the result to a total RMS aberration drawn uniformly at random from $[0 \text{ rad}, 1 \text{ rad}]$. This improved reconstruction performance significantly. We implement the neural networks using Flux.jl\cite{Innes2018} with integrated GPU support using CUDA.jl\cite{besard2018juliagpu,besard2019prototyping}. We measure a root-mean-squared error over the test set of $9.19 \times 10^{-2}$ rad, and measure linearity curves shown in Figure~\ref{fig:nn_sim}.

\subsection{Gerchberg-Saxton algorithm}
\label{sec:GS}

The Gerchberg-Saxton (G-S) algorithm is a phase retrieval technique that iteratively propagates a phase screen forwards and backwards through an accurate numerical model of a wavefront sensor. It contrasts with the neural network reconstructor and the interaction matrix approach by being model-driven rather than data-driven. It is an appropriate choice in this case because our model for the PL is relatively simple, involving only one matrix multiplication and norm-squared calculation. Further, rather than assuming the phase is well approximated by one or two terms in a Taylor series, as is respectively done by linear reconstruction or by the quadratic reconstructor explored in Lin et al. (2022)\cite{Lin22}, G-S is able to carry out wavefront reconstruction with fidelity that is equivalent to an arbitrarily-high-order expansion of the phase. Overall, although its practical implementation is likely to be slow due to the large number of iterations and high computational load, G-S is of interest to us as a representative of more tractable model-driven phase retrieval algorithms for the PL. The G-S algorithm is currently being used with the pyramid wavefront sensor on SEAL to find the initial best-flat position on startup.\cite{VincentGS}

The G-S algorithm starts with an assumed phase screen and propagates it forward to a wavefront sensor output. It then replaces the intensity of this WFS output with the measured intensity, while preserving the phase, and back-propagates this to the pupil plane to recover a new phase screen. This process repeats until convergence. In the context of the PL, the forward pass involves a pupil-to-focal plane propagation (with the \textit{hcipy.FraunhoferPropagator} object) and the decomposition of the resulting FP electric field in the lantern basis. In the backward pass, we reconstruct the input electric field by taking a linear combination of the input basis weighted by the lantern outputs, and back-propagate to the pupil plane. 

\begin{figure}
    \centering
    \includegraphics[width=\textwidth]{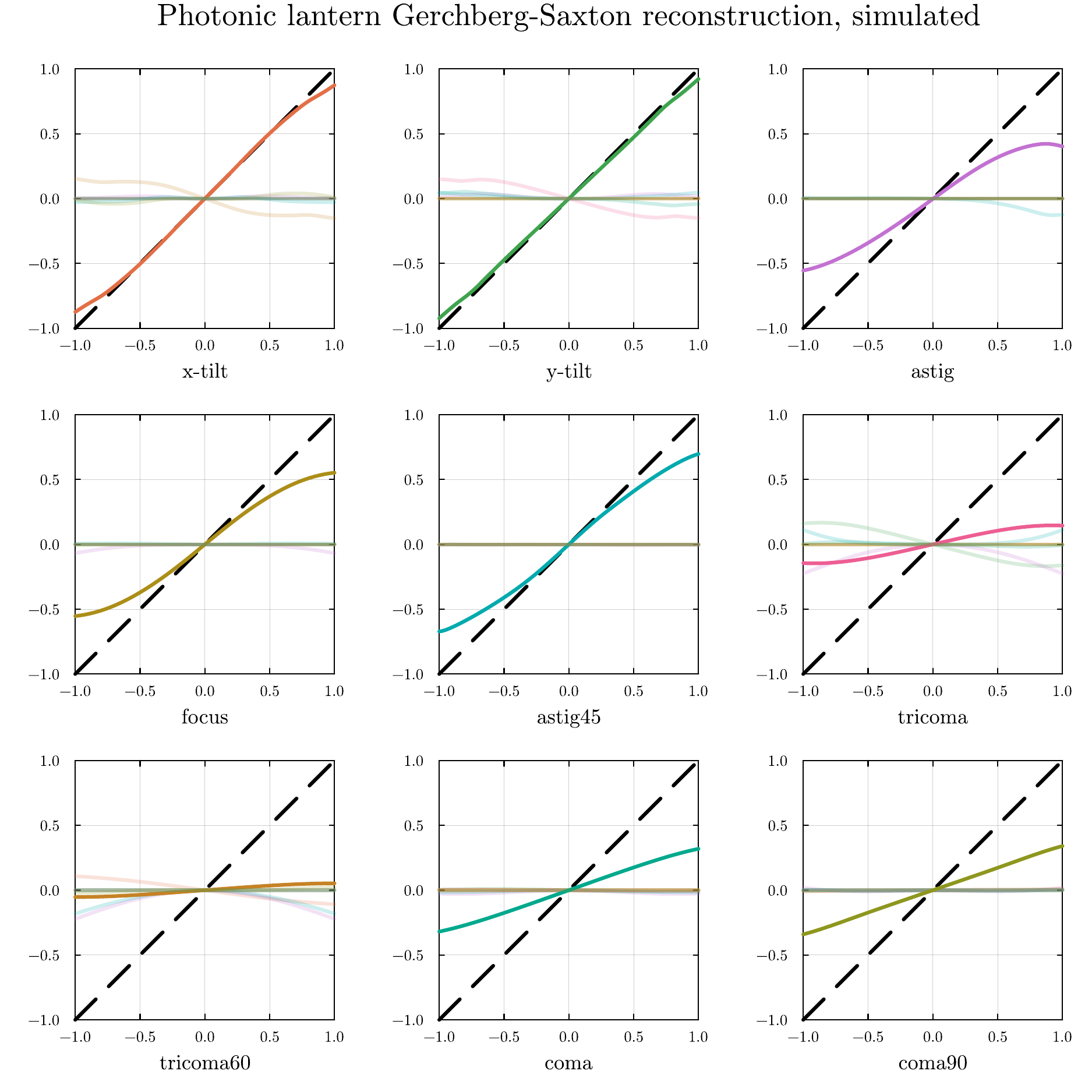}
    \caption{Linearity curves from the Gerchberg-Saxton algorithm.}
    \label{fig:GSlinearity}
\end{figure}

Figure~\ref{fig:GSlinearity} shows the linearity curves from this method. We observe minimal modal cross-talk and wide dynamic ranges, but increasing divergence from linearity with increasing spatial order. This is likely due to the effect of spatial filtering: the PL is more sensitive to aberrations resulting in a focal-plane image that mostly causes changes within the PL input in the focal plane. Further, G-S is invariant to overall intensity offsets, making it less sensitive to shifts in overall illumination of the output ports that are caused by high-amplitude aberrations. Other model-based reconstructors may avoid this issue.

In practice, G-S needs knowledge of the lantern's propagation matrix: a linear map from a basis of the supported electric fields at the input (i.e. those that excite a response in the lantern) to the corresponding complex amplitude on the fundamental mode on each lantern output. This is an $N \times N$ complex-valued matrix. We know this exactly in simulation -- our process of back-propagating from each single mode fiber in turn is the necessary condition for this matrix to be the identity -- and in the following subsection, we describe an algorithm to find this matrix experimentally.

\subsection{Empirical characterization algorithm}

We assume we have a light source, a wavefront affector (deformable mirror or spatial light modulator) and an $N$-port photonic lantern with a camera. We also assume we have a model of the optics before the PL. We are only able to apply some range of phases and record the corresponding PL image. We are interested in the linear system between the FP electric field and the PL output electric field, which we describe in a matrix $A_{N \times N}$, but we cannot probe it directly, as we can only control the input field via pupil-plane phases and can only observe the output in intensity. That is, we observe responses $y$ to input phases $\phi$:

\begin{align}
    y = \abs{A\mathcal{F}(P e^{i\phi})}^2
\end{align}

where $\mathcal{F}$ denotes a pupil-to-focal propagation and $P$ denotes the pupil amplitude function. We assume we can perfectly model $\mathcal{F}$ but cannot create arbitrary input electric fields: that is, the map $\phi \mapsto \mathcal{F}(P e^{i\phi})$ is injective but not surjective over the space of FP electric fields. Under these constraints, we can identify this matrix up to an overall phase difference, which does not affect intensity measurements. We achieve this by introducing phase diversity, recording the corresponding PL outputs, and solving a linear system per PL port that relates these outputs to the phase components of the propagation matrix.

\begin{figure}
    \centering
    \includegraphics[width=\textwidth]{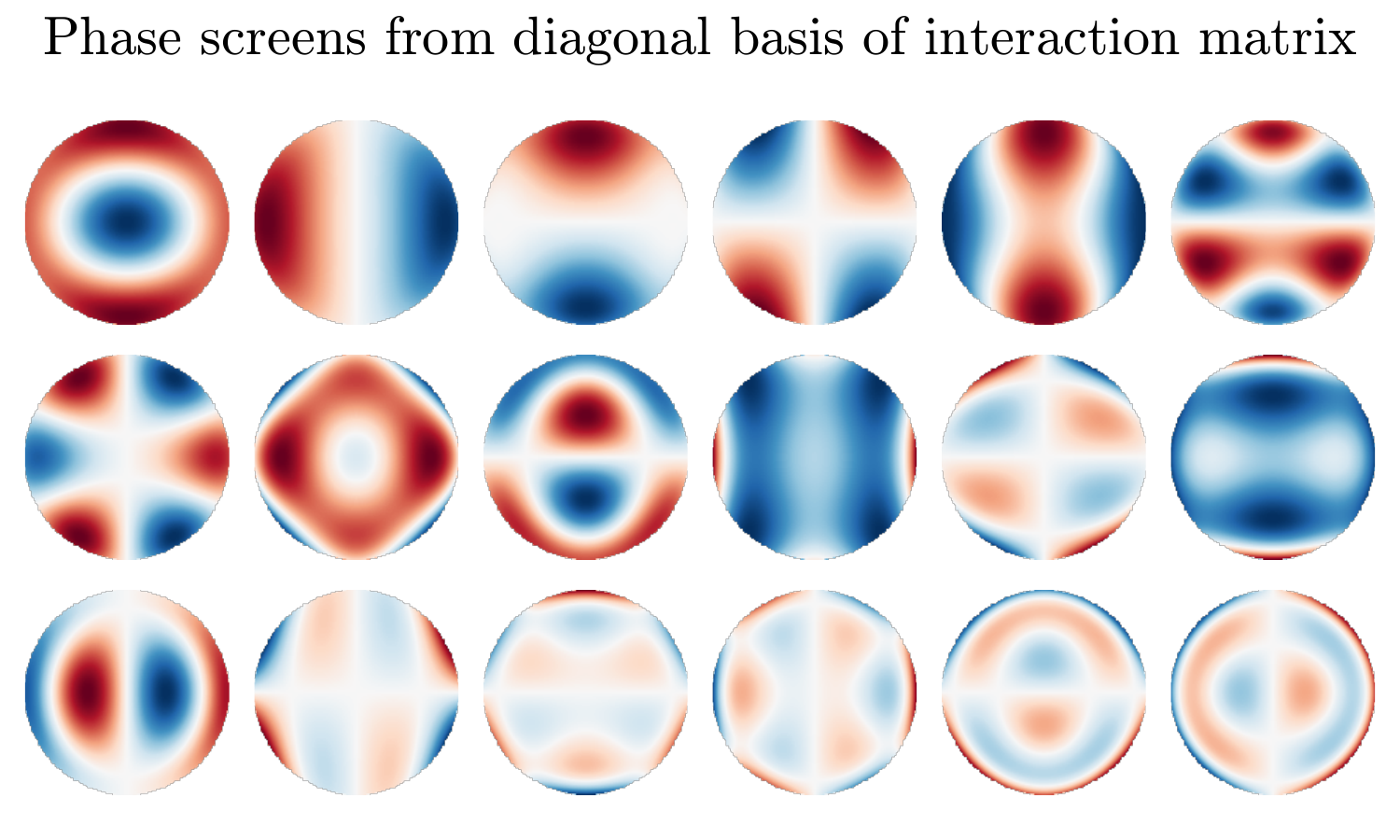}
    \caption{Phase screens in the diagonal basis of the interaction matrix. The characterization algorithm uses the focal-plane electric fields corresponding to these as a basis.}
    \label{fig:diagonal_phase_screens}
\end{figure}

We first define the basis for the FP electric field (henceforth the ``fitting basis"). Lin et al. (2022)\cite{Lin22} use the LP modes as the fitting basis, but we make a different choice for empirical characterization, as we are not guaranteed that there exists a phase screen that produces each LP mode at the lantern entrance. Instead, we find a basis by computing an interaction matrix over $N-1$ Zernike modes and taking its singular value decomposition, $IM = USV^\intercal$. The columns of $V^\intercal$ represent phase screens in Zernike space; Figure~\ref{fig:diagonal_phase_screens} shows these phase screens for our simulation. We use our model to propagate these to the focal plane. Taken together with the electric field for a flat wavefront, these form the desired fitting basis. The resulting transformation matrix has rank $N$, and therefore our propagation matrix $A$ will also be full-rank; alternate choices of phase screen sets have been found to have lower rank.

The square roots of the PL outputs for the fitting basis form the columns of the matrix $A^{\text{abs}}$, the element-wise absolute value of $A$. If we could observe the PL outputs' electric field, this would be sufficient; however, our result is off by some unknown phase factor:

\begin{align}
    \label{eq:recon}
    A_{pk} = A^{\text{abs}}_{pk} \exp(i \Phi_{pk})
\end{align}

where $p$ indexes the output port and $k$ indexes the element of the input basis. It remains to identify $\Phi_{pk}$.

We do this by applying $N(N-1)$ ``query" phase screens to the wavefront affector. These should be randomly distributed within the achievable range of the wavefront affector, such that all modes on the PL experience roughly their full range of variance. For each query, we compute the corresponding FP electric field and its projection in the fitting basis; any orthogonal component to the fitting basis has been experimentally shown to only minimally affect any lantern output. Suppose query $j$ has a fitting-basis electric field $f_j$; the response of the lantern in port $p$ is then given by

\begin{align}
    y_i = \abs{Af_j}^2 = \abs{\sum_{k=1}^N A_{pk} f_{kj}}^2 = \sum_{k=1}^N |A_{pk}|^2 |f_{kj}|^2 + 2 \sum_{l=1}^N \sum_{m=1}^{l-1} |A_{pl} f_{lj}| |A_{pm} f_{mj}| \cos(\mathrm{arg}(A_{pl} f_{lj}) - \mathrm{arg}(A_{pm} f_{mi}))
\end{align}

This is made up of a diagonal term and a cross term. Since we have identified $A^{\text{abs}}$, we know the diagonal term exactly, which lets us define the cross-term coefficients $c_j$:

\begin{align}
    c_j = y_j - \sum_{k=1}^N (A^{\text{abs}}_{pk})^2 |f_{kj}|^2 = 2 \sum_{l=1}^N \sum_{m=1}^{l-1} |A_{pl} f_{lj}| |A_{pm} f_{mj}| \cos(\mathrm{arg}(A_{pl} f_{lj}) - \mathrm{arg}(A_{pm} f_{mj}))
\end{align}

We further expand the cosine term in terms of the $\Phi_{pk}$s. For convenience, let $\phi_{ab} = \mathrm{arg} f_{ab}$.

\begin{align}
    \begin{split}
    c_j &= 2 \sum_{l=1}^N \sum_{m=1}^{l-1} |A_{pl} f_{lj}| |A_{pm} f_{mj}| \cos(\Phi_{pl} + \phi_{lj} - \Phi_{pm} + \phi_{mj})\\ 
    &= 2 \sum_{l=1}^N \sum_{m=1}^{l-1} \abs{A_{pl}} \abs{A_{pm}} \abs{f_{lj}} \abs{f_{mj}} [\cos(\phi_{lj} - \phi_{mj}) \cos(\Phi_{pl} - \Phi_{pm}) - \sin(\phi_{lj} - \phi_{mj}) \sin(\Phi_{pl} - \Phi_{pm})]
    \end{split}
\end{align}

This defines a linear system over $\left\{\cos(\Phi_{pl} - \Phi_{pm}), \sin(\Phi_{pl} - \Phi_{pm})\right\}$. Concretely, we define a matrix of cross-term coefficients $X^p$ whose terms are given by 

\begin{align}
    \begin{split}
        X^p_{2n,i} &= 2\abs{A_{pl}} \abs{A_{pm}} \abs{f_{lj}} \abs{f_{mj}} \cos(\phi_{lj} - \phi_{mj})\\
        X^p_{2n+1,j} &= -2\abs{A_{pl}} \abs{A_{pm}} \abs{f_{lj}} \abs{f_{mj}} \sin(\phi_{lj} - \phi_{mi})
    \end{split}
\end{align}

where $(l, m) \iff n$ defines an index convention that lets us describe the set of all pairs. We find the cosines and sines of phase differences $d^p$ for each port by solving the linear system $X^p d^p = c^p$ for each port $p$ (where $c^p$ consists of the $p$th component of $c_j$ for all $j$). 

From the system of cosines and sines, we reconstruct phases by taking

\begin{align}
    \theta(\sin\theta, \cos\theta) = \mathrm{sign}(\sin\theta) \times \cos^{-1}(\cos\theta)
\end{align}

and we use the system of phase differences to find precise phases. Since we are subject to an overall phase shift, we can take each $\Phi_{p0} = 0$ and take our measurement of $\Phi_{pk} - \Phi_{p0}$ as $\Phi_{pk}$; in the presence of read noise, it is possible to make this more robust by averaging over all the phase difference components involving $\Phi_{pk}$. Finally, we reconstruct $A$ by applying Equation~\ref{eq:recon}.

This procedure yields a reliable, physics-based photonic lantern forward model for use in wavefront reconstruction algorithms and for science applications. This enables further development of model-based reconstructors as well as optimization of lantern design for combined wavefront sensing and imaging performance.

\subsection{Comparison to modulated pyramid wavefront sensor}

\begin{figure}
    \centering\includegraphics[width=\textwidth]{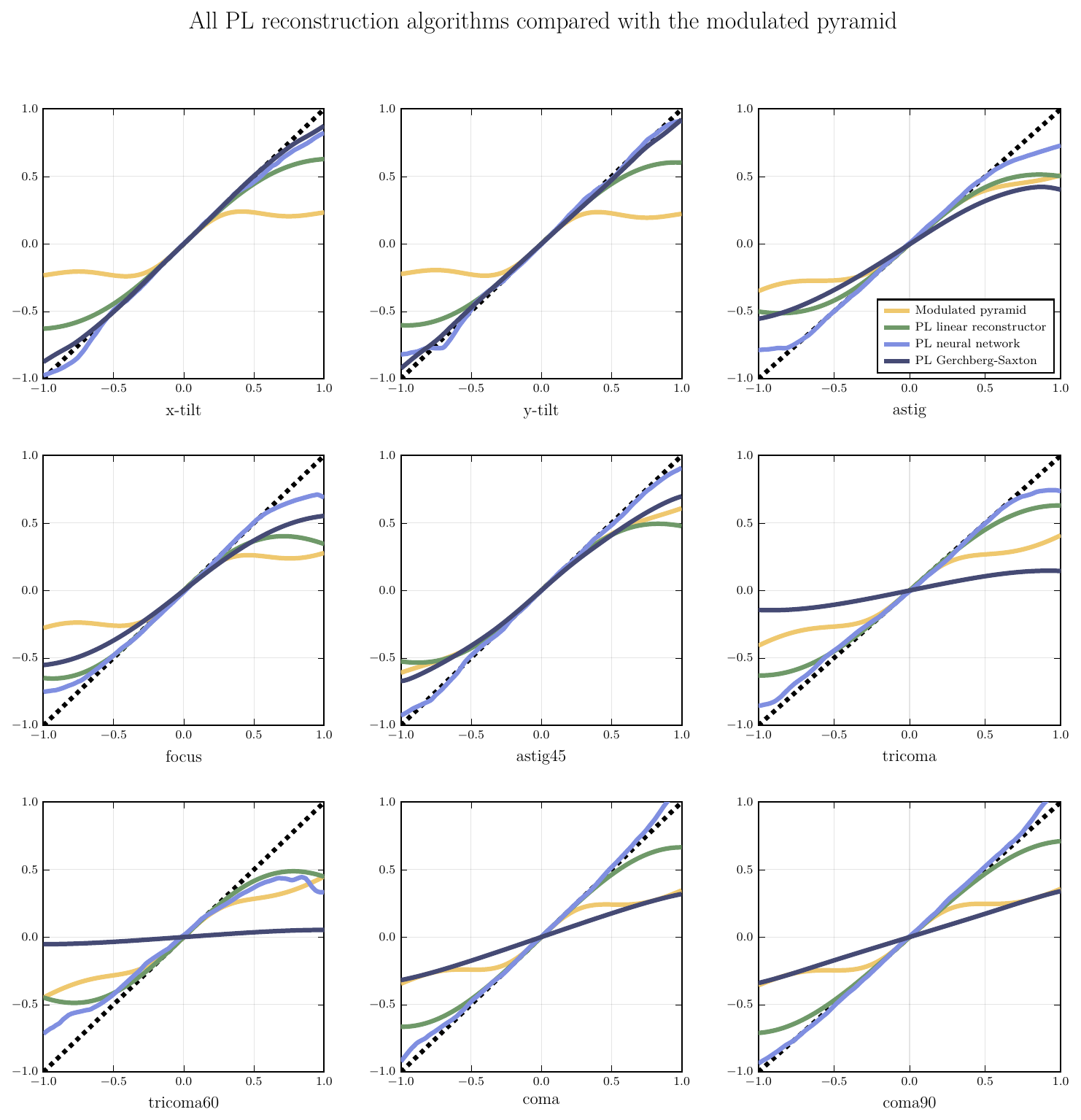}
    \caption{Superimposed linearity curves (only in the injected mode) for the modulated pyramid WFS and all three reconstruction techniques using the photonic lantern.}
    \label{fig:all_recon}
\end{figure}

Figure~\ref{fig:all_recon} shows the linearity curves for the three reconstruction techniques compared with the linearity curve for linear reconstruction with the modulated pyramid wavefront sensor.

\begin{figure}
    \centering
    \includegraphics[width=\textwidth]{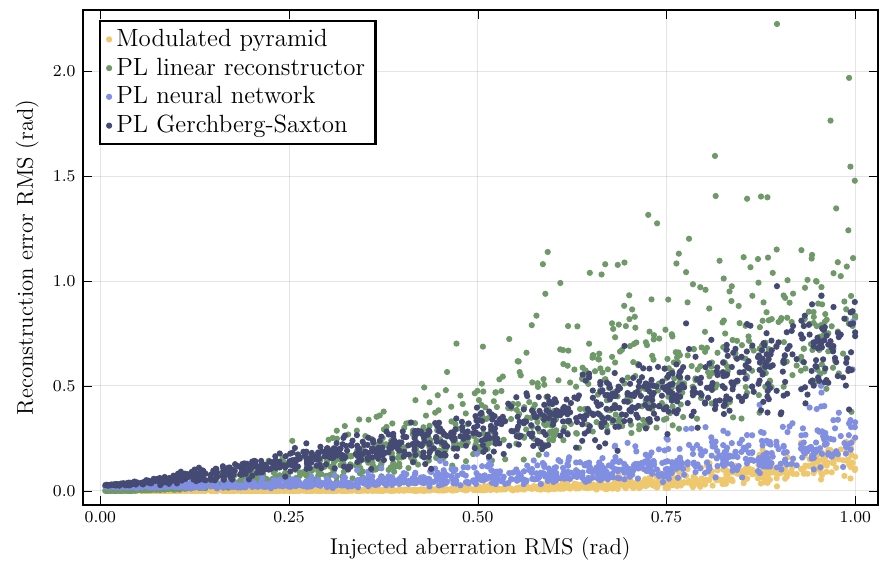}
    \caption{Reconstruction error for random combinations of the first 9 Zernike modes. We measure the RMS of the input phase screen and the RMS of the residual (reconstructed - input) phase screen.}
    \label{fig:nonlinear_recon}
\end{figure}

Figure~\ref{fig:nonlinear_recon} shows the average reconstruction error for arbitrary mixed modes as a function of the RMS of the injected phase screen, using the modulated pyramid and all three PL reconstructors. The neural network performs well at all injected aberrations. We note that the pyramid shows smaller linear ranges than the PL with all reconstruction methods in x-tilt/y-tilt/focus, but shows better performance than the PL on mixed-mode phase screens. This demonstrates the inherent nonlinearity of the PL and suggests that further development of nonlinear reconstructors could improve performance even further.

The Gerchberg-Saxton algorithm outperforms the linear reconstructor but performs worse than the neural network. The performance of the Gerchberg-Saxton method is likely limited by the spatial filtering issues discussed in Section~\ref{sec:GS}; however, we note that the reconstructed wavefronts are still mostly positively correlated with the ground truth, and so this method is still likely to result in favorable closed-loop AO performance. 

\section{MULTI WAVEFRONT SENSOR EXPERIMENTS}
\label{sec:results}

\subsection{Multi-wavefront sensor control architecture}

Our proposed multi-wavefront sensor architecture is defined similarly to Gerard et al. (2023)\cite{BenMultiWFS}. To ensure the two wavefront sensors do not reconstruct overlapping components of turbulence and result in a DM command that ``double-counts" aberrations, we apply temporal filters to the outputs of the wavefront sensors, based on an autoregressive framework first proposed by Gavel et al. (2014)\cite{Gavel2014} for woofer-tweeter control. The wavefront after the DM, as reconstructed by each wavefront sensor, is split into a low temporal frequency component used by the photonic lantern and a high temporal frequency component used by the modulated pyramid WFS. Based on a user-defined cutoff frequency $f_{\text{cutoff}}$, fixed to 30 Hz in this work, we split a signal $c_n$ into components $L_n, H_n$; assuming both wavefront sensors reconstruct the same signal within their control regime, we have $c_n = L_n + H_n$.

\begin{subequations}
\begin{align}
    \alpha &= \exp\left(-2\pi \frac{f_\text{cutoff}}{f_\text{loop}}\right)\label{eq:cutoff}\\
    H_n &= \alpha H_{n-1} + \alpha (c_n - c_{n-1})\\
    L_n &= \alpha L_{n-1} + (1 - \alpha) c_n\label{eq:lf}
\end{align}
\end{subequations}

Note that this is slightly different from Gerard et al. (2023): $\alpha$ has an additional factor of $2\pi$ in Equation~\ref{eq:cutoff} and the index on the $c$ term has changed from $n-1$ to $n$ in Equation~\ref{eq:lf}. These changes ensure that the true frequency cutoff lies close to $f_\text{cutoff}$ and that the full signal is reconstructed in ideal conditions.

These filters admit the following frequency representations:

\begin{subequations}
\begin{align}
    F_H(\omega) &= \frac{\alpha (1 - e^{-i\omega})}{1 - \alpha e^{-i\omega}}\\
    F_L(\omega) &= \frac{1 - \alpha}{1 - \alpha e^{-i\omega}}
\end{align}
\end{subequations}

\begin{figure}[!htbp]
    \centering
    \includegraphics[width=\textwidth]{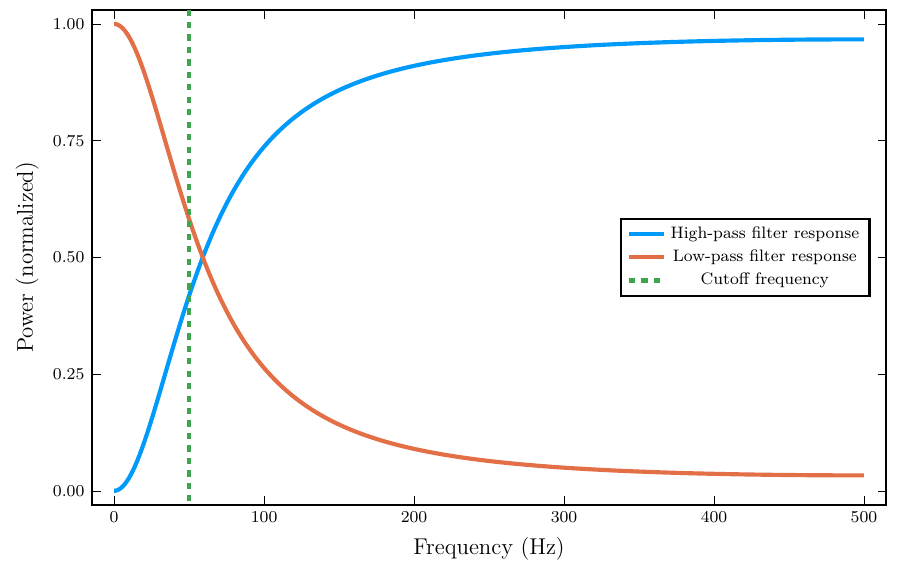}
    \caption{The power (squared magnitude of frequency response) of the high- and low-pass filtered wavefront sensor signals.}
    \label{fig:filter_freqres}
\end{figure}

Figure~\ref{fig:filter_freqres} shows the relative amount of power in both filtered signals as a function of frequency. In simulation, we can safely use the fact that the signal is split without any double-counting or lost components, but in real AO systems there may be relative frame delays between the two wavefront sensors that result in an imperfect overall reconstruction. This could be addressed by defining sharper filters by using higher autoregressive orders; for this work, it is sufficient to filter signals using the method shown here.

\begin{figure}[!htbp]
    \centering
    \begin{tikzpicture}[auto, node distance=3cm,>=latex', text width=2.2cm, align=center]
    \node [name=disturbance] {Atmospheric disturbance};
    \node [right of=disturbance, color=blue, node distance=5.5cm] (firststage) {Single-WFS configuration};
    \node [right of=firststage, color=red, opacity=0.75, node distance=4cm] (secondstage) {Double-WFS configuration};
    \node [sum, below of=disturbance, node distance=2cm, text width=0.5cm] (dm) {+};
    \node [draw, rectangle, below of=dm, node distance=2cm, minimum height=2em, minimum width=2em, text width=0.1cm, color=gray] (beamsplitter) {};
    \node [input, below right=1em and 1em of beamsplitter.center, anchor=center] (beamsplitterbr) {};
    \node [input, above left=1em and 1em of beamsplitter.center, anchor=center] (beamsplittertl) {};
    \node [input, below of=beamsplitter, node distance=2cm] (turn) {};
    \node [name=beamsplittertext, left of=beamsplitter, node distance=1.0cm] {Beam\\splitter};
    \node [block, minimum height=5em] (dmblock) at (dm) {};
    \node [above right=1.5em and 2em of dmblock.center, anchor=center] (dmtext) {DM};
    \node [block, right of=turn] (pl) {Photonic lantern};
    \node [block, right of=pl] (plrecon) {Lantern reconstructor};
    \node [block, right of=plrecon] (lpf) {Low pass filter};
    \node [block, right of=beamsplitter] (pywfs) {Pyramid WFS};
    \node [block, right of=pywfs,] (pyrecon) {Pyramid reconstructor};
    \node [input, right of=pyrecon, node distance=1.5cm] (pyswitch) {};
    \node [block, right of=pyswitch, node distance=1.5cm] (hpf) {High pass \\ filter};
    \node [input, above of=pyswitch, node distance=0.9cm] (turn4) {};
    \node [input, right of=hpf, node distance=1.4cm] (turn6) {};
    \node [input, right of=hpf, node distance=2.5cm] (hpfjoin) {};
    \node [input, above of=hpfjoin, node distance=0.9cm] (hpfjoinref) {};
    \node [input, left of=hpfjoinref, node distance=0.6325cm] (turn7) {};
    \node[input, below of=turn6, node distance=2cm] (lpfleftjoin) {};
    \node [input, below of=hpfjoin, node distance=2cm] (lpfjoin) {};
    \node [input, below of=turn7, node distance=2cm] (turn8) {};
    \node [sum, right of=hpf, text width=0.5cm, node distance=3.5cm] (sum2) {+};
    \node [input, right of=lpf, node distance=3.5cm] (turn2) {};
    \node [input, above of=sum2, node distance=2cm] (turn3) {};
    \node [gain, left of=turn3, shape border rotate=-180, text width=0.8cm, node distance=2.5cm] (gain) {Gain};
    \node [block, left of=gain, node distance=4.5cm] (integ) {Integrator};

    \filldraw (lpfjoin) circle (1pt);
    \filldraw (lpfleftjoin) circle (1pt);
    \filldraw (hpfjoin) circle (1pt);
    \filldraw (turn7) circle (1pt);
    \filldraw (turn6) circle (1pt);
    \filldraw (turn8) circle (1pt);

    \path[draw,->] (disturbance) -- (dm);
    \path[draw,->] (dm) -- (beamsplitter.center);
    \path[draw,-,color=gray] (beamsplittertl) -- (beamsplitterbr);
    \path[draw,->] (beamsplitter.center) -- (pywfs);
    \path[draw,->] (pywfs) -- (pyrecon);
    \path[draw,->] (pyrecon) -- (hpf);
    \path[draw,->] (hpfjoin) -- (sum2);
    \path[draw,-] (hpf) -- (turn6);
    \path[draw,-] (pyswitch) -- (turn4) -- (turn7);
    \path[draw,-] (beamsplitter.center) -- (turn);
    \path[draw,->] (turn) -- (pl);
    \path[draw,->] (pl) -- (plrecon);
    \path[draw,->] (plrecon) -- (lpf);
    \path[draw,-] (lpf) -- (lpfleftjoin);
    \path[draw,-] (lpfjoin) -- (turn2);
    \path[draw,->] (turn2) -- (sum2);
    \path[draw,-] (sum2) -- (turn3);
    \path[draw,->] (turn3) -- (gain);
    \path[draw,->] (gain) -- (integ);
    \path[draw,->] (integ) -- (dm) node[above, pos=0.95] {--};
    \path[draw,-,color=blue] (turn7) -- (hpfjoin);
    \path[->,color=red,opacity=0.75] (turn7) edge[bend right] (turn6);
    \path[draw,-,color=blue] (turn8) -- (lpfjoin);
    \path[->,color=red,opacity=0.75] (turn8) edge[bend right] (lpfleftjoin);
\end{tikzpicture}
    \caption{The multi-WFS control architecture used in this work. Loops are first closed in the single-WFS configuration without temporal filtering, then temporal filters are applied when the photonic lantern is added.}
    \label{fig:blockdiag}
\end{figure}
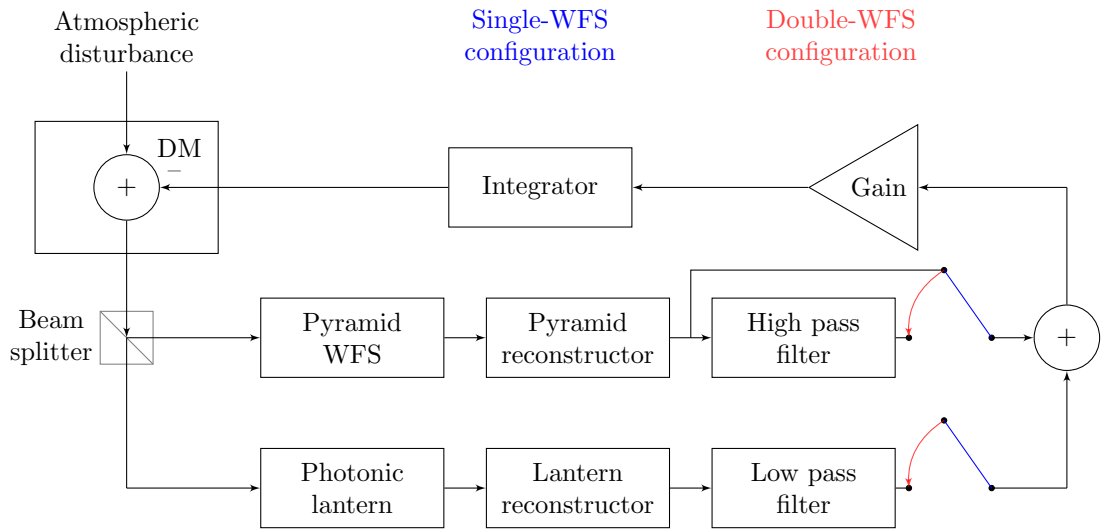

The control architecture we propose is shown in Figure~\ref{fig:blockdiag}. Future work will further investigate this control architecture in simulation and laboratory demonstrations, including temporal filter design and its impact on loop stability.

\subsection{Adaptive optics performance}

\begin{figure}
    \centering
    \includegraphics[width=\textwidth]{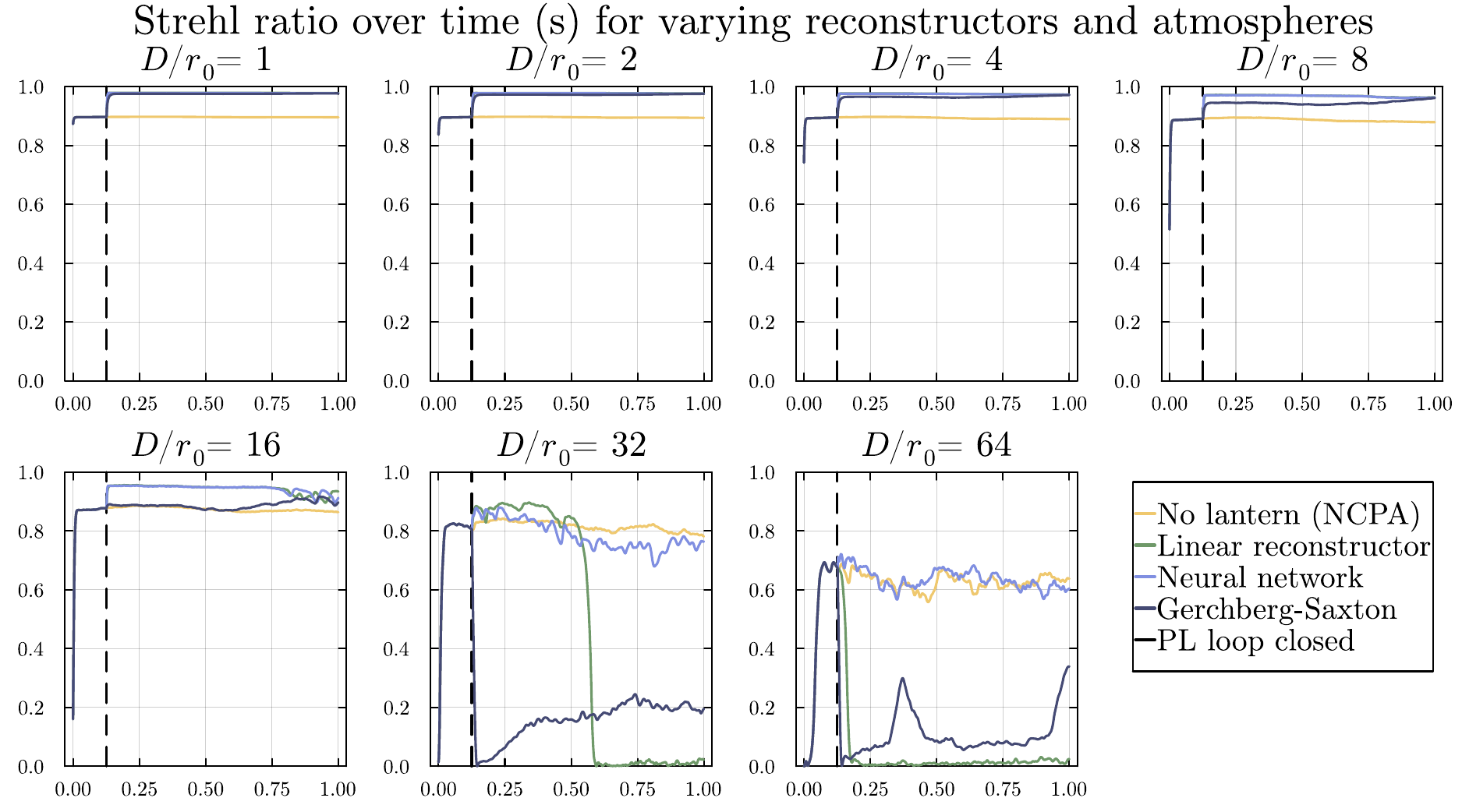}
    \caption{Strehl ratio over time for various reconstruction strategies. For each value of $D/r_0$, the same realization of atmospheric turbulence is used in each configuration. We close the PL loop after 100 frames (0.125s) in each case.}
    \label{fig:strehl_over_time}
\end{figure}

We simulate an AO loop using an integrator with a gain of 0.3, a leak of 0.999, and a loop rate of 800 Hz. Figure~\ref{fig:strehl_over_time} shows the Strehl ratio over time for four different configurations: a single-WFS system with no PL, and each of the three reconstructors discussed above. We introduce a static non-common-path aberration (a focus term of 0.3 rad) that cannot be seen by the pyramid, and note that it is corrected at low $D/r_0$. At $D/r_0 = 32$ or 64, it is initially compensated for, before other sources of inaccuracy begin to dominate.

\begin{figure}
    \centering
    \includegraphics{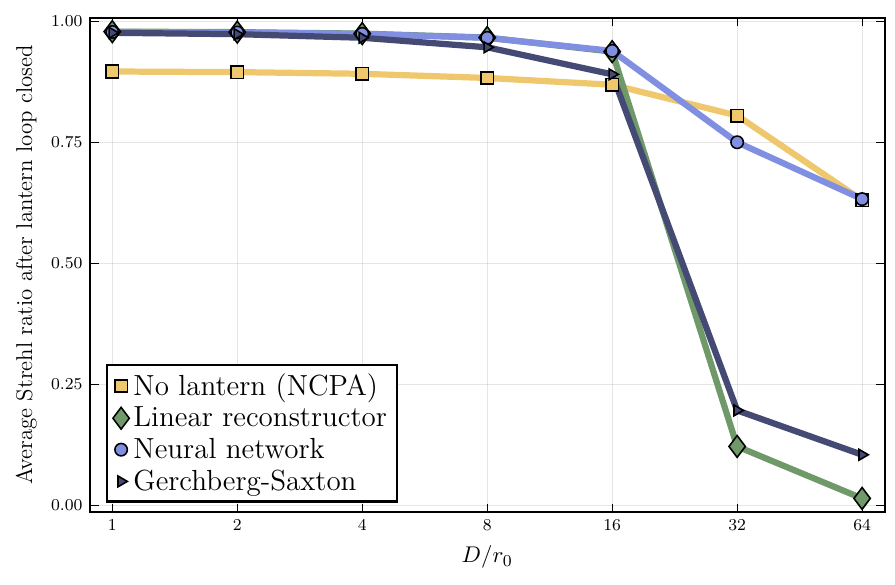}
    \caption{The eventual Strehl ratios achieved by each configuration as a function of $D/r_0$.}
    \label{fig:eventual_strehl}
\end{figure}

Figure~\ref{fig:eventual_strehl} shows the eventual Strehl ratios achieved in each configuration, calculated by averaging the Strehl ratio delivered over the last 400 frames (0.5s) of each simulation. We note that up to moderate aberrations, all reconstructors outperform the single-WFS case by accounting for the NCPA. The Gerchberg-Saxton algorithm begins underperforming relative to the other reconstructors at relatively small aberrations, which we attribute to low sensitivity as a result of its invariance to overall intensities. However, reconstructor error and/or loop instability causes all the multi-WFS cases to deliver Strehl ratios lower than the single-WFS case at high $D/r_0$ values, and the Gerchberg-Saxton algorithm slightly outperforms the linear reconstructor in this case, likely due to its large dynamic range. Further improvements in reconstruction algorithms and in the multi-WFS control architecture will help to address these cases.

\section{CONCLUSION}
\label{sec:conclusion}

We have demonstrated closed-loop control on a photonic lantern using the SEAL testbed, and have presented results from three wavefront reconstruction methods in simulation. Further, we have presented two new techniques towards making photonic lanterns work in real AO loops: an algorithm to constrain a lantern's propagation matrix empirically for use in scientific imaging applications and model-based reconstruction algorithms, and simulations of closed-loop control in a multi-wavefront sensor single-conjugate AO system for each reconstructor. This will enhance our ability to make optimal use of photonic lanterns for wavefront sensing in the lab and on sky. 

Future work will include lab testing of these reconstructors and various control architectures. Access to a photonic lantern that can be operated at its design wavelength will enable further testing of both model-driven and data-driven wavefront reconstruction algorithms, as well as extensions of the control scheme that can be optimized for the characteristics of each wavefront sensor being used.

\acknowledgments     
This work was funded in part by The Heising-Simons Foundation Grant 2020-1822. A.S. thanks Jonathan Lin for assistance with running \textit{lightbeam} and Parth Nobel for consultations while developing the empirical characterization algorithm. This document number is LLNL-CONF-865347. This work was performed under the auspices of the U.S. Department of Energy by Lawrence Livermore National Laboratory under Contract DE-AC52-07NA27344. 

\bibliography{report} 
\bibliographystyle{spiebib} 

\end{document}